\documentclass[a4paper]{jpconf}
\usepackage{graphicx, amsmath, mathtools}
\usepackage{color}

\begin{document}

\title{An iterative aggregation and disaggregation approach to the calculation of steady state distributions of continuous processes}

\author{Katja Biswas}

\address{Department of Physics, University of North Georgia, US}

\ead{Katja.Biswas@ung.edu}

\begin{abstract}
A mapping of the process on a continuous configuration space to the symbolic representation of the motion on a discrete state space will be combined with an iterative aggregation and disaggregation (IAD) procedure to obtain steady state distributions of the process.   
The IAD speeds up the convergence to the unit eigenvector, which is the steady state distribution, by forming smaller aggregated matrices whose unit eigenvector solutions are used to refine approximations of the steady state vector until convergence is reached. This method works very efficiently and can be used together with distributed or parallel computing methods to obtain high resolution images of the steady state distribution of complex atomistic or energy landscape type problems. The method is illustrated in two numerical examples. In the first example the transition matrix is assumed to be known. The second example represents an overdamped Brownian motion process subject to a dichotomously changing external potential.
\end{abstract}

\section{Introduction}

In recent years the projection of continuous processes onto discrete state space models has gained popularity. This is motivated by the fact that for systems consisting of a large number of atoms (such as bio-polymers) the dynamical process is often too complicated to be understood intuitively from single long time trajectories  and in many cases also computationally uneconomical. The latter particularly holds true for processes that evolve on disparate time and length scales. One approach to overcome these problems consists of building coarse discrete models representing the dynamics of the system. They consist of a set of abstract states and are represented by a stochastic matrix whose entries represent the transition between the states. If chosen adequately, this matrix can lead to an intuitive interpretation of important features of the original process.  
In this context Markov state models (MSM) \cite{MSMbook,MSMbookSchuette} have gained widespread importance.
Constructed in a two-step procedure\cite{MSMbook}, the original dynamics is first projected onto a large set of microstates from which, following certain criteria, macrostates are build. For realistic bio-polymers, microstate models typically can consist of $10,000$ to $100,000$ states making them hard to tackle computationally. 
 
Here I will use the projection of a continuous process onto a discrete state space together with an iterative aggregation and disaggregation (IAD) procedure to obtain a high-resolution steady state distribution of the process.
 The iterative aggregation and disaggregation methods are a class of methods \cite{Steward2008, Pultarova1,Pultarova2,me} known for being able to deal very efficiently with large and sparse stochastic matrices \cite{google1,google2}. Iteratively, the IAD methods calculate the eigenvector pertaining to the unit eigenvalue, which is the steady state distribution. Using an approximation of the steady state vector the stochastic matrix is aggregated to form a smaller matrix. The eigenvector problem is then solved on the smaller state space and, via a disaggregation step, its solution is used to refine the approximation of the steady state vector. 
In the projection of processes from the continuous phase- or configuration-\,space onto a discrete state space representation consisting of a large number of states, one will find that in majority of the cases a state will be connected to only a very limited number of other states. This leads to a large and sparse stochastic matrix \cite{me}, for which the unit eigenvector can be effectively computed using the IAD methods. Therefore, combining the IAD with a phase- or configuration-\,space mapping procedure may  
 provide a valuable way to deal with problems on the microstate level and may give useful information for macrostate clustering.

The article is structured as follows. In section 2 the method will be introduced, starting with the mapping procedure from the continuous process onto the discrete state space, followed by the IAD method. In section 3, the method will be illustrated at two numerical examples. 3.1 shows a simple case for which a $(5\times 5)$ transition matrix is known. This didactical example will be used to illustrate the steps in the iterative calculation of the steady state distribution. In section 3.2, the method will be shown at the Brownian motion process of a particle subject to a one dimensional potential which changes its shape with a dichotomous correlation in time. In this example the transition matrix P will be obtained using distributed computing and compared with those obtained from a full simulation.

\section{Method}

\begin{figure}[t!]
\centering
\begin{minipage}{0.48\linewidth}
\includegraphics[width=3.0in]{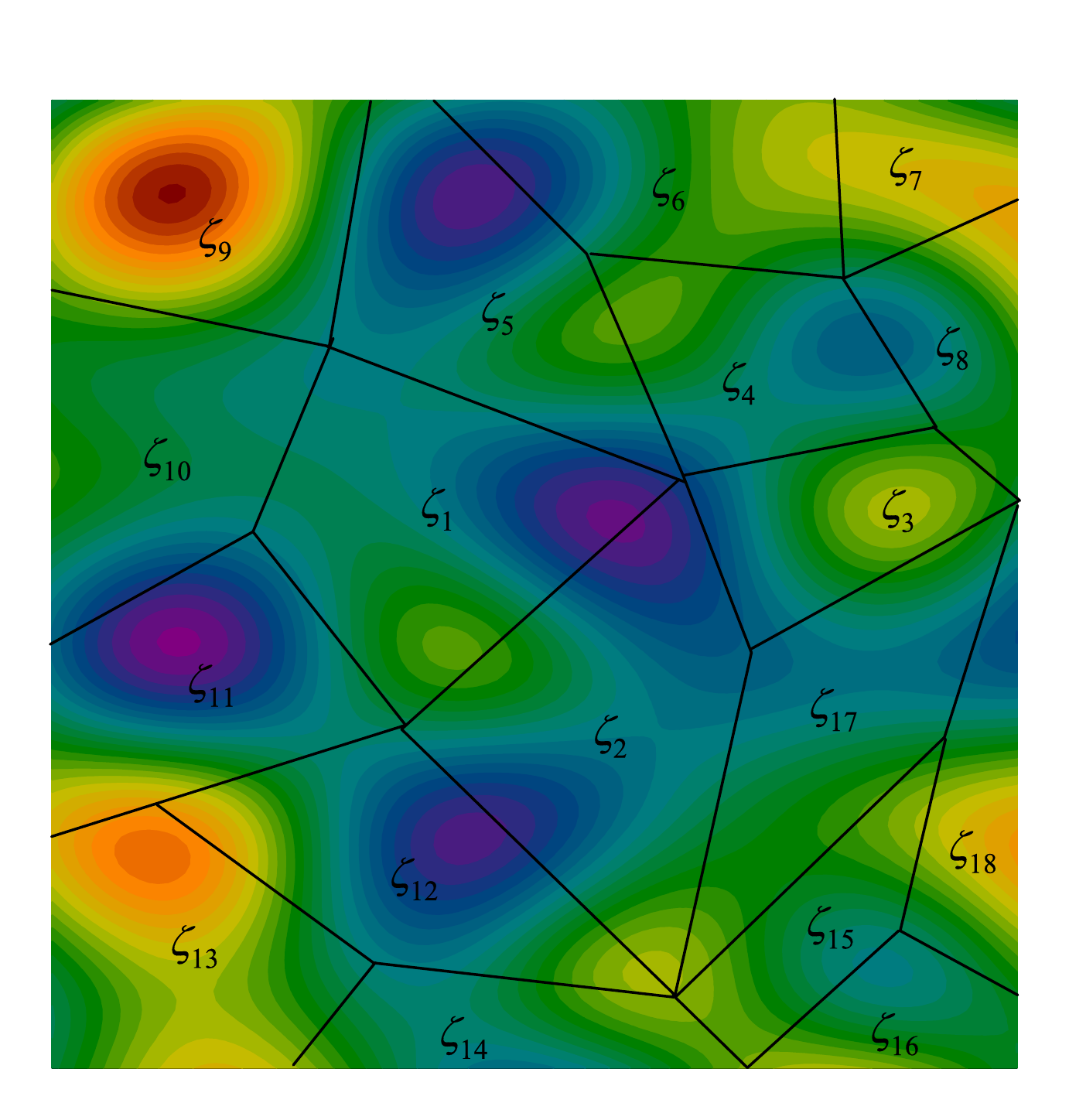}
\caption{\label{methfig1}Illustration of intervals on an arbitrary 2-dim energylandscape.}
\end{minipage}
\begin{minipage}{0.02\linewidth}
~
\end{minipage}
\begin{minipage}{0.48\linewidth}
\includegraphics[width=2.5in]{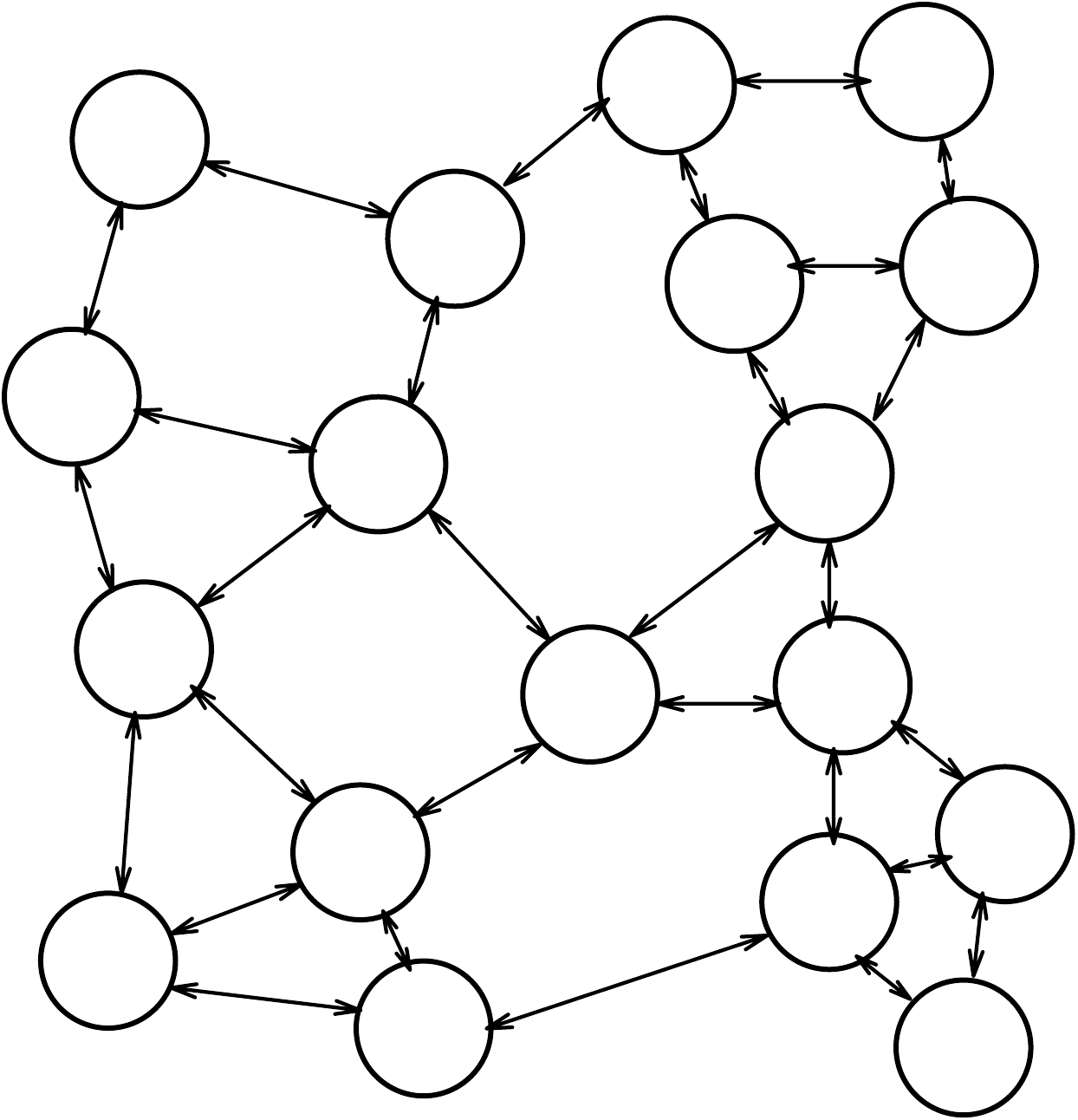}
\caption{\label{methodfig2}A discrete representation of the dynamics of a system.}
\end{minipage}
\end{figure}

Dynamical processes which are described by continuous phase- or configuration-\,space trajectories can be represented by the probabilities of transition between an abstract set of states. For this purpose, the accessible phase- or configuration-\,space needs to be projected onto a set of discrete states which I will label $i$, here $i=1,\ldots,m$ and $m$ is the total number of states. This can be done by dividing the phase- or configuration-\,space into $m$ non-overlapping intervals labeled $\zeta_i$ [see fig.\,(\ref{methfig1})]. A trajectory $\Gamma_l$ is then said to be found in state $i$ at time $t$ if 
$h_{\zeta_i}[\Gamma_l(t)]=1$, where
\begin{equation}
h_{\zeta_i}[\Gamma_l(t)]=\left\{
\begin{array}{c l}
 1&~~~  \text{, if}\quad\Gamma_l(t)\in\zeta_i\\
 0 &~~~\text{, otherwise} 
\end{array}\right. 
\end{equation} 
is the indicator function. This is used to measure the conditional probabilities $p(j|i)$ representing the probabilities at which a system that resides in state $i$ at time $t$ will have made a transition from state $i$ to state $j$ within a single time step $\mathrm{d}t$. They are given by\footnote{Eqn.\,(\ref{mappingeqn}) is an expression of $p(j,t+\mathrm{d}t|i,t)p(i,t)=p(j,t+\mathrm{d}t;i,t)$, where $p(j,t+\mathrm{d}t;i,t)$ is the joined probability of finding the system at time $t$ in state $i$ and at time $t+\mathrm{d}t$ in state $j$. More details concerning the notation of the mapping procedure can be found in \cite{meJphysA11,meJphysA13}.}
\begin{equation}
p(j|i)=\left\{
\begin{array}{l l}
 \frac{ \langle h_{\zeta_j}[\Gamma_l(t+\mathrm{d}t)] h_{\zeta_i}[\Gamma_l(t)]\rangle}{\langle h_{\zeta_i}[\Gamma_l(t)]\rangle}
&~~~  \text{, if}\quad\langle h_{\zeta_i}[\Gamma_l(t)]\rangle \neq 0\\
 0 &~~~\text{, otherwise} 
\end{array}\right. \>,
\label{mappingeqn}\end{equation}
where the angular brackets $\langle \ldots \rangle$ denote the time and ensemble average.
 They can be used to set up the transition matrix $\mathbf{P}$ whose elements, the transition probabilities $p_{ij}$, are defined as
\begin{equation}
p_{ij}=\left\{
\begin{array}{l l}
 p(j|i)&~~~  \text{, if}\quad j\neq i\\
 1-\sum_{j\neq i} p(j|i) &~~~\text{, if}\quad j = i
\end{array}\right. \>.
\label{transitionprobabilities}\end{equation}
The $p_{ij}$, with $j\neq i$, denote the probabilities of the system making a transition from state $i$ to $j$ within a single time step $\mathrm{d}t$. The elements $p_{ii}$ denote the probability of a system that resides in state $\zeta_i$ to stay in the same state during the time step $\mathrm{d}t$. The transition matrix is then given as
\begin{equation}
\mathbf{P}=
\begin{pmatrix}
p_{11} & p_{12} & p_{13} & p_{14}& \cdots & p_{1m} \\
p_{21} & p_{22} & p_{23} & p_{24} &\cdots & p_{2m}\\
p_{31} & p_{32} & p_{33} & p_{34} &\cdots & p_{3m}\\
p_{41} & p_{42} & p_{43} & p_{44} &\cdots & p_{4m}\\
\vdots & \vdots & \vdots & \vdots  & \ddots & \vdots \\
p_{m1} & p_{m2} & p_{m3} & p_{m4} &\cdots & p_{mm} 
\end{pmatrix} \>.
\end{equation}

\begin{figure}[t!]
\begin{center}
\includegraphics[width=2.5in]{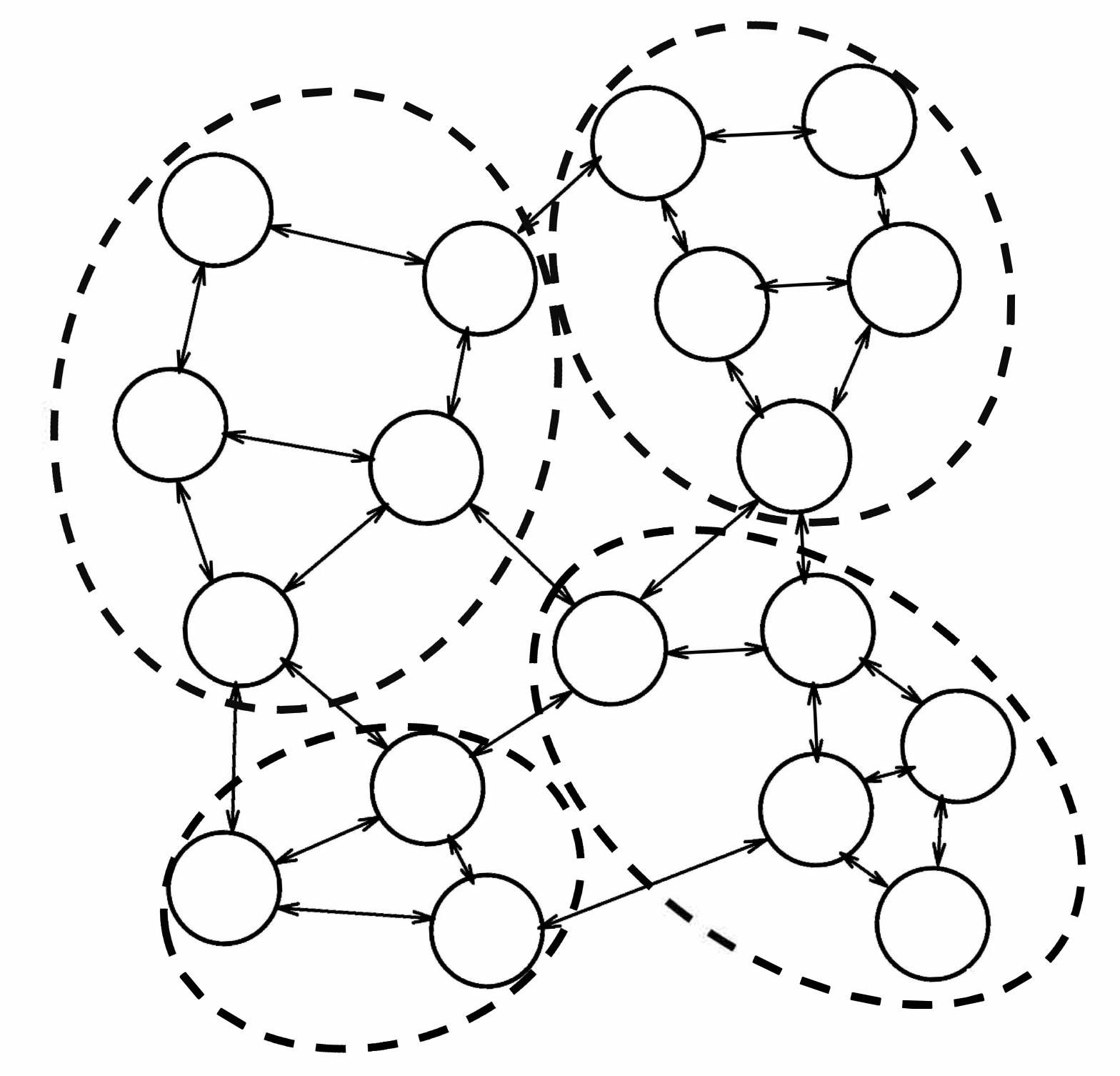}
\caption{\label{methodaggregates}Illustration of the partitioning of microstates. Note, that in this step the transition probabilities on the microstate level are preserved. }
\end{center}
\end{figure}

For a set of $m$ discrete states $i$ the matrix $\mathbf{P}$ has dimension $(m\times m )$. See fig.\,(\ref{methodfig2}) for a visualization.
 Particularly for partitions of the original continuous phase- or state-\,space which lead to a very large number of states, finding the steady state distribution can be a very arduous task. The steady state distribution is given by the eigenvector corresponding to the unit eigenvalue and corresponds to a situation in which the probabilities of occupation of the individual states are not changing with respect to time. In the following the steady state distribution will be obtained using an iterative aggregation and disaggregation procedure.\\
In order to apply the procedure, the matrix $\mathbf{P}$ has to be partitioned into block matrices. Roughly, this can be interpreted as outlining macrostates, but with the difference that the information about the transitions between the smaller states is preserved at this step [compare to fig.\,(\ref{methodaggregates})], i.e.
\[
\mathbf{P}=
\begin{pmatrix}
\begin{bmatrix}
p_{11} & p_{12} & p_{13}\\   
p_{21} & p_{22} & p_{23}\\ 
p_{31} & p_{32} & p_{33} 
\end{bmatrix} &
\begin{bmatrix}
 p_{14}& \cdots \\
p_{24} &\cdots\\
 p_{34} &\cdots 
\end{bmatrix} &
\cdots&
\begin{bmatrix}
p_{1(m-1)} & p_{1m} \\
p_{2(m-1)} & p_{2m}\\
p_{3(m-1)}& p_{3m}
\end{bmatrix}\\
\vdots & \vdots & \ddots & \vdots\\ 
\begin{bmatrix}
\vdots & \vdots & \vdots\\
p_{m1} & p_{m2} & p_{m3} 
\end{bmatrix}&
\begin{bmatrix}
\vdots &\vdots \\
 p_{m4} &\cdots 
\end{bmatrix}&
\cdots &
\begin{bmatrix}
\vdots & \vdots\\
p_{m(m-1)}& p_{mm}
\end{bmatrix}
\end{pmatrix} 
=
\begin{pmatrix}
\mathbf{P}_{11} & \mathbf{P}_{12} & \cdots & \mathbf{P}_{1\ell} \\
\mathbf{P}_{21} & \mathbf{P}_{22} & \cdots & \mathbf{P}_{2\ell}\\
\vdots & \vdots  & \ddots & \vdots \\
\mathbf{P}_{\ell 1} & \mathbf{P}_{\ell 2} & \cdots & \mathbf{P}_{\ell\ell}\\
\end{pmatrix} \>.
\]
The $\mathbf{P}_{ij}$ are block matrices and have the dimension $(m_i\times n_j)$, where $\sum_{i=1}^{\ell} m_i=\sum_{j=1}^{\ell} n_j = m$. Together with an approximation of the steady state eigenvector the block matrices are
 aggregated to form entries of a matrix $\mathbf{Q}$, which is used to solve the eigenvector problem on a smaller state space. The solution on the smaller state space is then disaggregated to improve the approximation of the steady state vector on the microstate space. The procedure is as follows:
\begin{enumerate}
\item\label{step1}{Choose the initial approximation of the steady state vector $\mathbf{x}^0=(x^0_1,x^0_2,\ldots ,x^0_m)$  and let $k=1$.}
\item\label{step2}{Compute the vectors $\mathbf{\hat{x}}_i^{(k-1)}$ as
\begin{equation}
\mathbf{\hat{x}}_i^{(k-1)}=\frac{\mathbf{x}_i^{(k-1)}}{\left \lVert \mathbf{x}_i^{(k-1)}\right\rVert_1} \>,
\notag\end{equation}
 for $i=1,\ldots ,\ell$, $\left \lVert \mathbf{x}_i^{(k-1)}\right\rVert_1=\mathbf{x}_i^{(k-1)}\mathbf{e}$ and $\mathbf{e}$ is a vector with all elements equal to one.  }
\item\label{step3}{Obtain the aggregated matrix $\mathbf{Q}^{(k-1)}$ whose elements $q^{(k-1)}_{ij}$ are given by
\begin{equation}
\mathbf{q}^{(k-1)}_{ij}=\mathbf{\hat{x}}_i^{(k-1)}\mathbf{P}_{ij}\mathbf{e} \>,
\notag\end{equation}
for $i,j=1,\ldots,\ell$.\footnote{Note, while the matrix $\mathbf{Q}$ represents the aggregated matrix, its entries are not to be confused with macrostate transition probabilities. They represent an aggregation of the iterative procedure $\mathbf{x}^{(k)}=\mathbf{x}^{(k-1)}\mathbf{P}$.}}
\item\label{step4}{Find the eigenvectors $\mathbf{w}^{(k-1)}=(w_1,w_2,\ldots ,w_\ell)$ corresponding to the unit eigenvalue of matrix $\mathbf{Q}^{(k-1)}$, i.e. solve the problem
\begin{equation}
\mathbf{w}^{(k-1)}\mathbf{Q}^{(k-1)}=\mathbf{w}^{(k-1)}
\notag\end{equation}
and
\begin{equation}
\sum_{i=1}^\ell w_i^{(k-1)} =1 \>.
\notag\end{equation}}
\item\label{step5}{Set up the diagonal matrix $\mathbf{Z}^{(k-1)}$ defined as
\begin{equation}
\mathbf{Z}^{(k-1)}=\text{diag}\left\{ \frac{\mathbf{w}_1^{(k-1)}}{\lVert \mathbf{x}_1^{(k-1)}\rVert_1} \mathbf{I}_{1}, 
 \frac{\mathbf{w}_2^{(k-1)}}{\lVert \mathbf{x}_2^{(k-1)}\rVert_1} \mathbf{I}_{2},
\ldots ,
 \frac{\mathbf{w}_\ell^{(k-1)}}{\lVert \mathbf{x}_\ell^{(k-1)}\rVert_1} \mathbf{I}_{\ell}
\right\}\>,
\notag\end{equation}
where $\mathbf{I}_{j}$ are identity matrices of dimension $(n_j\times n_j)$. }
\item\label{step6}{Improve the approximation to the steady state vector via
\begin{equation}
\mathbf{x}^{(k)}=\mathbf{x}^{(k-1)}\mathbf{Z}^{(k-1)}\mathbf{K} \>,
\notag\end{equation}
with
\begin{equation}
\sum_{i=1}^m x^{(k)}_i=1 \>.
\notag\end{equation}
}
\item\label{step7}{Check for convergence of the vector $\mathbf{x}^{(k)}$. This can be done by calculating $\eta=\lVert \mathbf{x}^{(k)}-\mathbf{x}^{(k-1)}\rVert$, where $\lVert \ldots \rVert$ is some norm. If $\eta< \epsilon$ the iteration procedure has reached its desired accuracy and can be exited, otherwise set $k=k+1$ and continue at step (\ref{step2}).} 
\end{enumerate}

In step (\ref{step6}) $\mathbf{K}=\mathbf{L}(\mathbf{D}-\mathbf{U})^{-1}$, where $\mathbf{L}$ and $\mathbf{U}$ are the lower triangular and the upper triangular matrix of the transition matrix $\mathbf{P}$, and $\mathbf{D}=\mathbf{I}-\text{diag}\{\mathbf{P}\}$ is a diagonal matrix with the elements given by $d_{ii}=1-p_{ii}$ for $i=1,\ldots ,m$. $\mathbf{I}$ is the identity matrix and has dimension $(m\times m)$. 
 $\mathbf{K}$ has to be calculated only once and does not need to be updated during the iteration procedure.
Further, convergence of the method may be improved by  implementing a pre- or post-smoothing procedure\,\cite{me}. In the pre-smoothing procedure the matrix $\mathbf{P}$ is multiplied by itself, i.e. $\mathbf{P}\rightarrow\mathbf{P}^{s+1}$, where $s$ is the number of pre-smoothing steps and $s=0$ corresponds to the method without pre-smoothing steps. The post-smoothing procedure is implemented by raising $\mathbf{K}$ to a power, i.e. $\mathbf{K}\rightarrow\mathbf{K}^{r+1}$, where $r$ denotes the number of post-smoothing steps and $r=0$ corresponds to the method without post-smoothing steps. 

\section{Numerical Examples}
In this section I will show two examples. In the first example the iterative aggregation and disaggregation method will be illustrated at the calculation of the steady state vector of a system for which the matrix $\mathbf{P}$ is known. This example is for didactical purpose, it will be used to illustrate steps (\ref{step1}) to (\ref{step7}) of the procedure. In the second example the method will be illustrated at particles undergoing Brownian motion over a one-dimensional fluctuating potential. The fluctuations represent a dichotomous change of the potential.  
\subsection{Didactic example with known transition matrix}

Let the matrix $\mathbf{P}$ be a $(5\times 5)$ matrix with the following entries
\[
\mathbf{P} =
\begin{pmatrix}
    0.4       &  0.3 & 0 & 0.15 & 0.15 \\
    0.2     & 0.5 & 0.2 & 0.1 & 0 \\
    0.1 & 0.1 & 0.6 & 0.05 & 0.15 \\
  0 & 0.2 & 0.4 & 0.2 & 0.2 \\
 0.2 & 0 & 0 & 0.1 & 0.7
\end{pmatrix} \>.
\]
Matrix $\mathbf{I}-\mathbf{P}$ decomposes into $\mathbf{L}$, $\mathbf{D}$ and $\mathbf{U}$ in the following way
\[
\begin{pmatrix}
    0       &  0 & 0 & 0 & 0 \\
    0.2     & 0 & 0 & 0 & 0 \\
    0.1 & 0.1 & 0 & 0 & 0 \\
  0 & 0.2 & 0.4 & 0 & 0 \\
 0.2 & 0 & 0 & 0.1 & 0
\end{pmatrix},
\begin{pmatrix}
    0.6  &  0 & 0 & 0 & 0 \\
    0 & 0.5 & 0 & 0 & 0 \\
    0 & 0 & 0.4 & 0 & 0 \\
    0 & 0 & 0 & 0.8 & 0 \\
    0 & 0 & 0 & 0 & 0.3
\end{pmatrix}
\text{and}
\begin{pmatrix}
    0  &  0.3 & 0 & 0.15 & 0.15 \\
    0  & 0 & 0.2 & 0.1 & 0 \\
    0 & 0 & 0 & 0.05 & 0.15 \\
  0 & 0 & 0 & 0 & 0.2 \\
 0 & 0 & 0 & 0 & 0
\end{pmatrix} \>.
 \] 
These are the lower triangular, the diagonal and the upper triangular matrix (respectively) of the decomposition $\mathbf{I}-\mathbf{P}=\mathbf{D}-\mathbf{L}-\mathbf{U}$, where $\mathbf{I}$ is the identity matrix. This gives for $\mathbf{K}=\mathbf{L}\left(\mathbf{D}-\mathbf{U}\right)^{-1}$ 
\[
\begin{pmatrix}
0 & 0 & 0 & 0 & 0 \\
0.3333 & 0.2 & 0.1 & 0.0938 & 0.2792\\
0.1667 & 0.3 & 0.15 & 0.0781 & 0.2104 \\
0 & 0.4 & 1.2 & 0.125 & 0.6833 \\
0.3333 & 0.2 & 0.1 & 0.2188 & 0.3625
\end{pmatrix}\>.
\]
Do obtain the aggregated matrix $\mathbf{Q}$, the matrix $\mathbf{P}$ is partitioned into four submatrices $\mathbf{P}_{ij}$ (with $i,j=1,2$) in the following way
\[
\left( \begin{matrix}
   \begin{bmatrix}
      0.4 & 0.3 & 0\\
    0.2 & 0.5 & 0.2\\
    0.1 & 0.1 & 0.6\\
  \end{bmatrix} & \begin{bmatrix}
  0.15 & 0.15 \\
  0.1 & 0 \\
 0.05 & 0.15
\end{bmatrix}\\\!\rule{0in}{.28in}
\begin{bmatrix}
 0 & 0.2 & 0.4 \\
 0.2 & 0 & 0 \\
\end{bmatrix} & \begin{bmatrix}
 0.2 & 0.2 \\
 0.1 & 0.7
\end{bmatrix}
\end{matrix}\right)
=
\begin{pmatrix}
\mathbf{P}_{11} & \mathbf{P}_{12} \\
\mathbf{P}_{21} & \mathbf{P}_{22}
\end{pmatrix}\>.
\]
The iterative procedure is then initiated according to step (\ref{step1}) by choosing a starting approximation $\mathbf{x}^0$ to the steady state vector. For this purpose, let $\mathbf{x}^0=(0.2 ~~ 0.2 ~~ 0.2 ~~ 0.2 ~~ 0.2)$, which is then partitioned into subvectors compatible with the partitioning of matrix $\mathbf{P}$ into the submatrices $\mathbf{P}_{ij}$, i.e.
\[
\begin{pmatrix}
\begin{bmatrix} 
0.2 & 0.2 & 0.2
\end{bmatrix} & 
\begin{bmatrix} 
0.2 & 0.2 
\end{bmatrix}
\end{pmatrix} =
\begin{pmatrix}
\mathbf{x}^0_1 & \mathbf{x}^0_2
\end{pmatrix} \>.
\]
Normalizing each of the subvectors gives for step (\ref{step2}) 
\[
\begin{pmatrix}
\mathbf{\hat{x}}^0_1 & \mathbf{\hat{x}}^0_2
\end{pmatrix}=
\begin{pmatrix}
\begin{bmatrix}
\frac{1}{3} & \frac{1}{3} & \frac{1}{3}
\end{bmatrix} &
\begin{bmatrix}
\frac{1}{2} & \frac{1}{2}
\end{bmatrix}
\end{pmatrix}\>.
\] 
Note, the elements $\left(\mathbf{\hat{x}}^0_n\right)_{j_n}$ represent the approximation of the conditional probability of finding the system in one of the microstates $j_n$ given it is in the partition $n$ belonging to microstate $j_n$ (here $n=1,2$ denotes the partition and $j_1=1,\ldots,3$ and $j_2=4,5$ are the microstates). The aggregated matrix $\mathbf{Q}^0$ of step (\ref{step3}) then becomes
\[
\mathbf{Q}^0=
\begin{pmatrix}
0.8 & 0.2\\
0.4 & 0.6
\end{pmatrix}\>.
\]
Solving the eigenvector problem on the aggregated state space, i.e. step (\ref{step4}), gives the aggregated eigenvector  
\[
\mathbf{w}^0 = 
\begin{pmatrix}
\frac{2}{3} & \frac{1}{3}
\end{pmatrix} \>,
\]
and the diagonal matrix $\mathbf{Z}^0$ of step (\ref{step5}) becomes
\[
\mathbf{Z}^0=
\begin{pmatrix}
\frac{10}{9} & 0 & 0 & 0 & 0 \\
0 & \frac{10}{9} & 0 & 0 & 0 \\
0 & 0 & \frac{10}{9} & 0 & 0 \\
0 & 0 & 0 & \frac{5}{6} & 0 \\
0 & 0 & 0 & 0 & \frac{5}{6}
\end{pmatrix}\>.
\]
Note, here the entries in the diagonal matrix $\mathbf{Z}$ indicate that probability is transferred from the subvector $\mathbf{\hat{x}}^0_2$ to $\mathbf{\hat{x}}^0_1$.
The disaggregation step\,(\ref{step6}) gives for $\mathbf{x}^{(1)}$ after normalization
\[
\mathbf{x}^{(1)}=
\begin{pmatrix}
0.162 & 0.2052 & 0.2647 & 0.0928 & 0.2752
\end{pmatrix}\>.
\]
Implementing the Euclidian norm $\eta=\sqrt{\sum_i (x^{(k)}_i-x^{(k-1)}_i)^2}$ for step\,(\ref{step7}) gives
\[
\eta=0.15101 \>.
\]
To reach a convergence better than $\epsilon=10^{-5}$ the procedure has to be repeated eight more times, leading to 
\[
\mathbf{x}^{(9)}=
\begin{pmatrix}
0.1959 &0.2041 & 0.2123 &0.1102&0.2775
\end{pmatrix}\>.
\]
As comparison, implementing the standard iterative procedure $\mathbf{x}^{(k)}=\mathbf{x}^{(k-1)}\mathbf{P}$ requires eighteen iterations to reach the same convergence.

\subsection{Computational Example - Fluctuating Barrier}

\begin{figure}[t!]
\centering
\begin{minipage}{0.48\textwidth}
\includegraphics[width=3.25in]{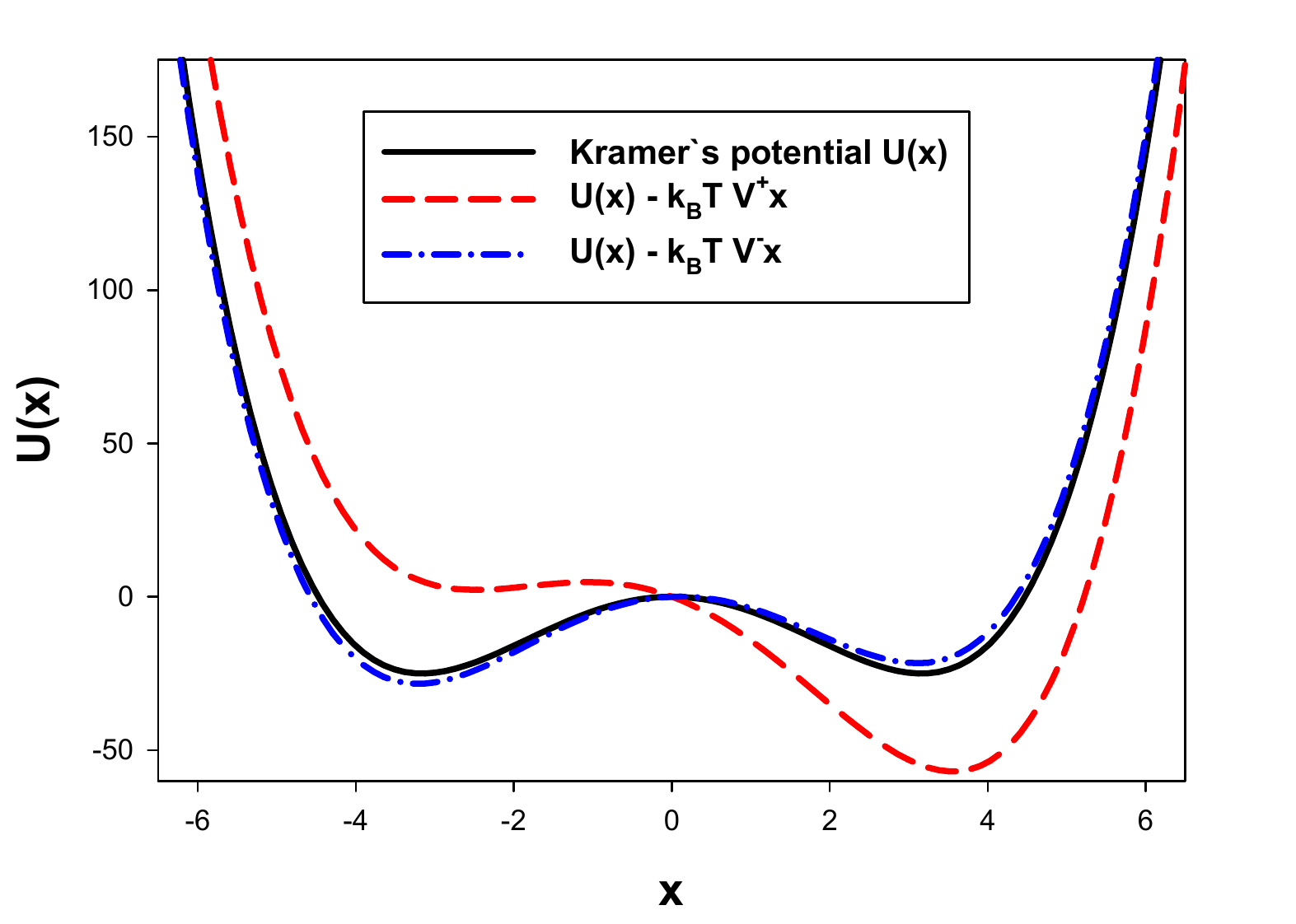}
\caption{\label{figdicho1}Plot of the effective potential in its $V^+$ (dashed line) and $V^-$ state (dash-dot line). The unaltered potential is drawn as solid line. }
\end{minipage}
\begin{minipage}{0.02\textwidth}
~
\end{minipage}
\begin{minipage}{0.48\textwidth}
\includegraphics[width=3.25in]{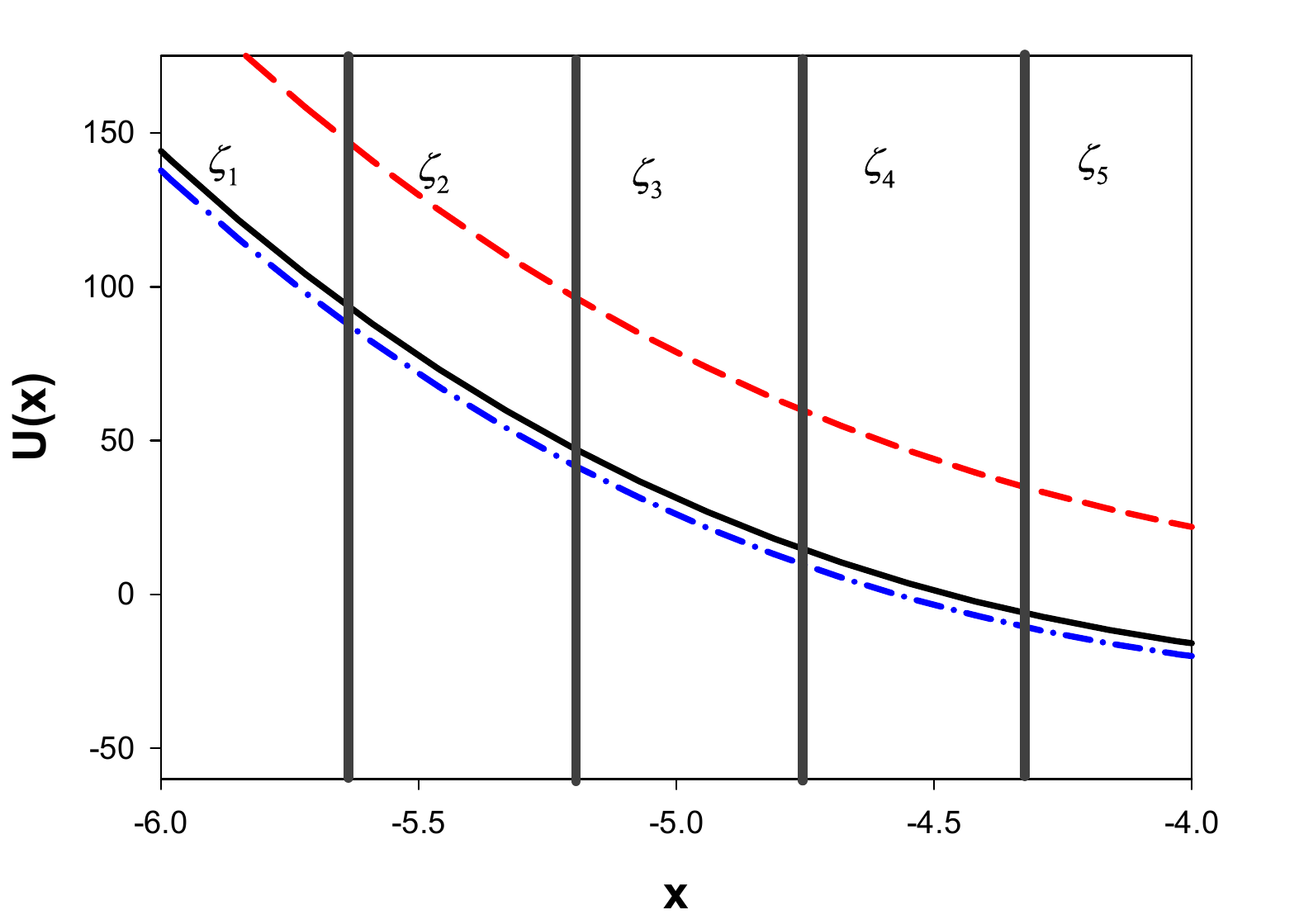}
\caption{\label{figdicho2} Illustration of the first five intervals $\zeta_1,\ldots,\zeta_5$ used to define the states. Note that these intervals were also used to define limited regions for distributed computing.}
\end{minipage}
\end{figure}

To illustrate the full procedure, a one dimensional example is chosen where the dynamics represents an overdamped Brownian motion subject to a dichotomously changing potential \cite{Armin}, i.e.
\begin{equation}
\dot{x}=-\nabla \left[ u(x) +V_{DN}(\tau)x\right] +\Gamma (\tau)  \>,
\label{dynamics}\end{equation}
where $\eta$ is the friction constant, $\tau=\frac{k_BT}{\eta}\,t$, $u(x)=U(x)/k_BT$ and $\Gamma(\tau)$ is a white noise given by the following properties
\[
\langle\Gamma (\tau)\rangle =0
\]
\[
\langle \Gamma (\tau) \Gamma (\tau')\rangle =2 \delta (\tau-\tau') \>.
\]
$V_{DN}$ represents a dichotomous noise alternating between the two fixed values
\[
V^{+}=\sqrt{\frac{D}{\tau_v}\left(\frac{1+\epsilon}{1-\epsilon}\right)}
\qquad\quad
\text{and}
\qquad\quad
V^{-}=-\sqrt{\frac{D}{\tau_v}\left(\frac{1-\epsilon}{1+\epsilon}\right)} \>,
\]
where $\tau_v$ is the correlation time of the dichotomous noise, $\epsilon$ a parameter introducing asymmetry and $\sqrt{D/\tau_v}$ the amplitude.
The switching between the $V^{+}$ and $V^{-}$ state of the extended potential is correlated in time following
\[
\langle V_{DN} (\tau)\rangle =0
\] 
and
\[
\langle  V_{DN} (\tau)V_{DN}(0)\rangle= \frac{D}{\tau_v}\exp \left(-\frac{\tau}{\tau_v}\right) \>.
\]
The potential $U(x)$ used in eq. (\ref{dynamics}) is Kramer`s potential, i.e.
\[
U(x)=-\frac{a}{2}x^2+\frac{b}{2}x^4 \>.
\]
The simulation is carried out using the predictor corrector algorithm for dynamics subject to dichotomous noise\,\cite{dynamicsIN} with a time step $\mathrm{d}t=3\times 10^{-4}$. The parameters used are $\eta=1$, $\epsilon=0.8$, $a=10$, $b=1$, $k_BT=0.15\Delta U$, where $\Delta U=a^2/4b$ is the height of the barrier between the two minima in Kramer`s potential, and $D/\tau_v=0.71$. A plot of the potential in its $V^{+}$ and $V^{-}$ state is shown in fig.\,(\ref{figdicho1}).
 
The entries $p_{ij}$ in the matrix $\mathbf{P}$ are obtained using distributed computing. For this purpose the interval $[-6.5, 6.5]$ is divided into $30$ intervals $\zeta_i$ of equal length [fig.\,(\ref{figdicho2}) shows the first five intervals] and simulations are carried out individually for each of the inner states in the following way. For each independent run 500 trajectories are initiated in the interval $\zeta_i$ for $i=2,\ldots,29$ and allowed to evolve in the combined interval $\zeta_{i-1}\bigcup\zeta_i\bigcup\zeta_{i+1}$, which allows to take into account local recrossing effects\footnote{Local in the sense that it takes the configuration-space history of the trajectory into account only up to the neighboring states, but not farther.}. Trajectories which attempt to leave the combined interval via a transition from $\zeta_{i-1}$ to $\zeta_{i-2}$ (or from $\zeta_{i+1}$ to $\zeta_{i+2}$) are reflected at the boundaries of the simulation region back into $\zeta_{i-1}$ (or $\zeta_{i+1}$ respectively). After sufficient initial run, necessary to remove the effects of the starting location of the trajectories on the measurement of the transition probabilities, the single time step conditional probabilities $p(i-1|i)$ and $p(i+1|i)$ (for $i=2,\ldots ,29$) of leaving state $\zeta_i$ to $\zeta_{i-1}$ and to $\zeta_{i+1}$ are measured using 10mio time steps. \\
 The conditional probabilities are then used to set up the transition matrix $\mathbf{P}$ using the transition probabilities $p_{ij}$ according to eq.\,(\ref{transitionprobabilities}). Note, during the simulation it is observed that some of the states are not reached from other states. These states represent high energy outer states of the potential energy, and are cut out of the matrix $\mathbf{P}$. This leads to tri-diagonal Markov-state matrices of dimension $(25\times 25)$  for the correlation times $\tau_v=3\times 10^{-3},\ldots,1\times 10^{-2}$ and $\tau_v=6\times 10^{-1},\ldots, 2.5\times 10^1$ and to matrices of dimension $(26\times 26)$ for $\tau_v=2.5\times 10^{-2},\ldots ,2.5\times 10^{-1}$ and $\tau_v=6\times 10^1,\ldots,6\times 10^{3}$.\\
For the case of 25 connected states, the matrix $\mathbf{P}$ is then partitioned into submatrices $\mathbf{P}_{ij}$ of dimension $(5\times 5)$, leading to an aggregated matrix $\mathbf{Q}$ of dimension $(5\times 5)$. For the case of $26$ connected states, the elements $p_{ij}$ with $i,j=1,\ldots ,15$ are partitioned into submatrices as before and the elements $p_{(26)j}$ with $j=1,\ldots, 26$ and $p_{i(26)}$ with $i=1,\ldots ,25$ are partitioned as $(1\times 5)$- and $(5\times 1)$-dimensional submatrices $\mathbf{P}_{6j}$ for $j=1,\ldots 5$ and $\mathbf{P}_{i6}$ for $i=1,\ldots 5$. The submatrix $\mathbf{P}_{66}$ is comprised of only one element, namely $p_{(26)(26)}$.   \\
The iterative procedure is started [i.e. step\,(\ref{step1})] with an equal distribution of the initial approximation $\mathbf{x}^0$, i.e. $x_i=1/25~(\forall ~ i=1,\ldots,25)$ or $x_i=1/26 ~(\forall ~ i=1,\ldots ,26)$ respectively, and stopped once the norm $\eta=||x^{(k)}-x^{(k-1)}||$ is smaller than $\epsilon=1\times 10^{-4}$. As norm the $L^2$ norm was chosen, i.e. the Euclidian norm where $\eta=\sqrt{\sum_i (x_i^{(k)}-x_i^{(k-1)})^2}$.

\begin{figure}[t!]
\centering
\begin{minipage}{0.48\textwidth}
\includegraphics[width=3.25in]{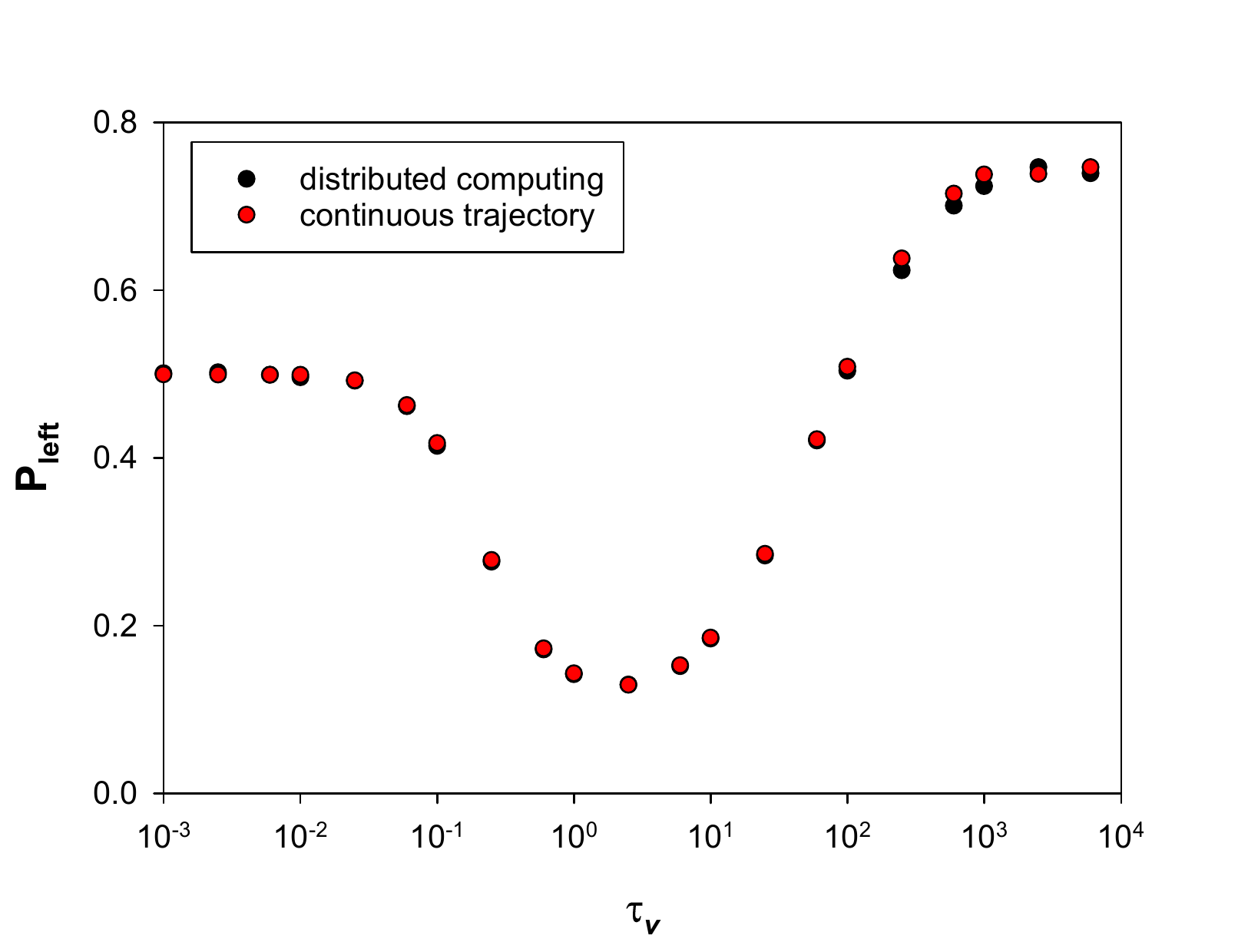}
\caption{\label{figdicho4}Steady state probability of occupation of the left potential well $\mathbf{P}_{\text{left}}$ for different switching times $\tau_v$. The transition matrices were obtained using distributed computing (black dots) and continuous trajectories (red dots). }
\end{minipage}
\begin{minipage}{0.02\textwidth}
~
\end{minipage}
\begin{minipage}{0.48\textwidth}
\includegraphics[width=3.25in]{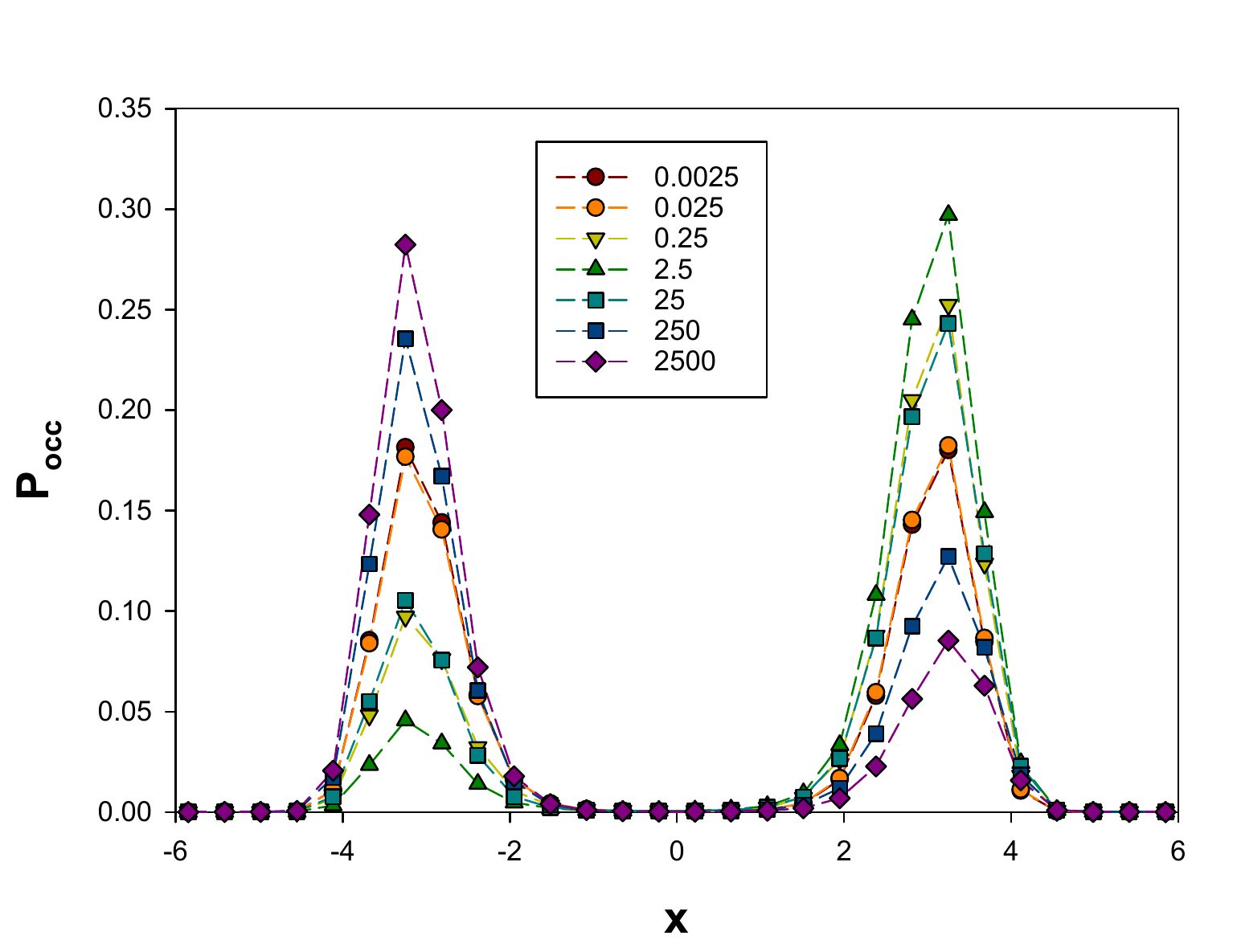}
\caption{\label{figdicho5} Steady state occupation of the states $\mathbf{x}$ for some of the switching times $\tau_v$.}
\end{minipage}
\end{figure}
 Figure (\ref{figdicho4}) shows for different dichotomous noise correlation times $\tau_v$ a plot of the probability of steady state occupation of the left potential well (i.e. the region in space for which $x\leq 0$). The results for the distributed computing are compared to those from a full simulation using continuous trajectories. As can be seen from the figure, for fast switching times the results are in agreement within the convergence bound of the iterative procedure. For slow switching times, i.e. $\tau_v>10^2$, the difference between the values is bigger. This could be an indication of the effect the distributed computing procedure could have on the accuracy of the transition probabilities. Overall, the results agree with those reported by other authors\,\cite{Armin}. A plot of the probabilities of occupation of the individual states is shown in fig.\,(\ref{figdicho5}) for some of the correlation times $\tau_v$ (numbers shown in the legend). It can be seen from fig.\,(\ref{figdicho5}) that at correlation times of $\tau_v=2.5\times 10^{-3}$ and $2.5\times 10^{-2}$ the left and right potential well are approximately equally occupied, whereas for slow correlation times states corresponding to the left potential well have higher occupancy than states within the right potential well. The minimum of the occupancy of the left potential well is found around $\tau_v=2.5$ [compare figure\,(\ref{figdicho4}) and (\ref{figdicho5})], indicating that at this dichotomous noise correlation time nonequilibrium kinetic focusing occurs.\\

\section{Summary}
In this paper a method is introduced which combines the mapping from a continuous process on the phase- or configuration-\,space to a stochastic matrix representing the probabilities of transitions between an abstract set of states, with that of an iterative aggregation and disaggregation procedure in an effort to effectively obtain the steady state distribution of the process. The method is shown on two examples. The first example is of didactical nature, highlighting the different steps in the iterative procedure, whereas the second example shows a Brownian motion process subject to a dichotomously changing potential. Here the transition probabilities of the stochastic matrix were obtained using distributed computing and the resulting probabilities of occupation were compared to those from full trajectory simulations for different rates of fluctuation. For correlation times of the dichotomous noise $\tau_v\leq10^2$  the results for the probability of occupation of the left potential well are in good agreement, within the convergence bounds, for both the procedures. Faster dichotomous noise correlation times show a larger difference between the steady state distributions, which possibly could indicate limitations of the distributed computing procedure and will be subject to further studies.\\
Note, the use of transition probabilities rather than direct trajectory information makes the method suitable for the use for large scale bio-physical processes. In this paper the method was implemented using a simple distributed computing procedure, where the transition probabilities were obtained from independent runs of different domains, but could also be combined with other simulation procedures such as trajectory parallelization\,\cite{Venturoli} or parallel replica dynamics\,\cite{Voter}.


\section{References}


\begin{thebibliography}{99}

\bibitem{MSMbook} G.R. Bowman and V.S. Pande and F. No\'e, 
{\it An Introduction to Markov State Models and Their Application to Long Timescale Molecular Simulation}, Springer, Dordrecht, 2014 (and references therein).

\bibitem{MSMbookSchuette}
  C. Sch{\"u}tte and M. Sarich,
  {\it Metastability and Markov State Models in Molecular Dynamics},
  Courant Lecture Notes, AMS,
  Providence,
  2013.


\bibitem{google1}
  I. Ipsen and S. Kirkland,
  {\it SIAM J. Matrix Anal. Appl.},
  {\bf 27},
  952 (2006).


\bibitem{google2}
  H. Sterck and T. Manteuffel and S. McCormic and Q. Nguyen and J. Ruge,
  {\it SIAM J. Sci. Comput.},
  {\bf 30},
   2235 (2008).

\bibitem{Steward2008}
  W.-L. Cao and W. Steward,
  {\it J. Assoc. Comput. Mach.},
  {\bf 32},
   702 (1985).

\bibitem{Pultarova1}
  I. Pultarov\'a and I. Marek,
  {\it Numer. Linear Algebra},
  {\bf 18},
  1051 (2011).

\bibitem{Pultarova2}
  I. Pultarov\'a and I. Marek,
  {\it J. Comput. Appl. Math.},
  {\bf 236},
  354 (2011).

\bibitem{me}
  K. Biswas,
  {\it Comput. Phys. Commun.},
  {\bf 191},
  25 (2015).

\bibitem{meJphysA11}
  K. Biswas and M. A. Novotny,
  {\it J. Phys. A: Math. Theor.},
  {\bf 44},
  345004 (2011).

\bibitem{meJphysA13}
  K. Biswas,
  {\it J. Phys. A: Math. Theor.},
  {\bf 46},
  145001 (2013).

\bibitem{Armin}
  L. Ponzoni and G.L. Celardo and F. Borgonovi and L. Kaplan and A. Kargol,
  {\it Phys. Rev. E},
  {\bf 87},
  052137 (2013).

\bibitem{dynamicsIN}
   Debashis Barik and Pulak Kumar Ghosh and Deb Shankar Ray,
  {\it J. Stat. Mech.},
  {\bf 03},
  P03010 (2006).

\bibitem{Venturoli}
   Eric Vanden-Eijnden and Maddalena Venturoli,
   {\it J. Chem. Phys.},
  {\bf 131},
  044120 (2009).

\bibitem{Voter}
   Danny Perez and Blas P. Uberuaga and Arthur F. Voter,
  {\it Comput. Mater. Sci.},
  {\bf 100},
  90 (2015).

\end{thebibliography}
\end{document}